\documentclass[%
    aps,
    prl,
    reprint,
    superscriptaddress,
    nolongbibliography,
    preprintnumbers,
    noeprint,
]{revtex4-2}

\usepackage[utf8]{inputenc}
\usepackage[T1]{fontenc}

\usepackage{newtxtext,newtxmath}

\usepackage{graphicx}
\usepackage[%
    allcolors=blue,
    colorlinks=true,
]{hyperref}
\usepackage{orcidlink}
\usepackage[%
    print-unity-mantissa=false,
    range-phrase=-,
    range-units=single,
]{siunitx}

\newcommand{\GRASP}{\affiliation{Institute for Gravitational and Subatomic Physics, Utrecht University, Princetonplein 1, 3584 CC Utrecht, Netherlands}}
\newcommand{\Nikhef}{\affiliation{Nikhef, Science Park 105, 1098 XG Amsterdam, Netherlands}}
\newcommand{\ITP}{\affiliation{Institute for Theoretical Physics, Utrecht University, Princetonplein 5, 3584 CC Utrecht, Netherlands}}

\newcommand{\bayesfactor}{\mathcal{B}_\nores^\res}
\newcommand{\bayesfactorthreshold}{\mathcal{B}_\text{th}}
\newcommand{\nores}{\text{no res}}

\newcommand{\res}{\text{res}}

\newcommand{\aap}{Astron.\ Astrophys.}
\newcommand{\apjss}{Astrophys.\ J.\ Suppl.\ Ser.}
\newcommand{\cqg}{Class.\ Quantum Gravity}
\newcommand{\jcap}{J.\ Cosmol.\ Astropart.\ Phys.}
\newcommand{\mnras}{Mon.\ Not.\ R.\ Astron.\ Soc.}
\newcommand{\natcom}{Nat.\ Commun.}

\newcommand{\prx}{Phys.\ Rev.\ X}

\begin{document}

\title{Detecting Tidal Resonances in Binary Neutron Stars}

\author{Fabian~Gittins~\orcidlink{0000-0002-9439-7701}}
\GRASP\Nikhef

\author{Harsh~Narola~\orcidlink{0000-0001-9161-7919}}
\GRASP\Nikhef

\author{Thibeau~Wouters~\orcidlink{0009-0006-2797-3808}}
\GRASP\Nikhef

\author{Peter~T.~H.~Pang~\orcidlink{0000-0001-7041-3239}}
\Nikhef\GRASP

\author{Tanja~Hinderer~\orcidlink{0000-0002-3394-6105}}
\ITP

\author{Chris~Van~Den~Broeck~\orcidlink{0000-0001-6800-4006}}
\GRASP\Nikhef

\date{October 19, 2026}

\begin{abstract}
    As a binary neutron star inspirals due to the emission of gravitational waves, the rising tidal frequency resonantly excites vibrational modes. These oscillations are seismological probes of the rich stellar interior, yet it remains to be established whether gravitational-wave interferometers can measure them. Here, we present the first fully Bayesian study of the capability of the Einstein Telescope to detect tidal resonances. We simulate one year of observations and analyse the 200 loudest signals. We find that the Einstein Telescope can identify resonant modes and is sensitive to gravitational-wave phase shifts as small as $\Delta \Phi \approx 0.03$ for favourable events. We further show that neglecting resonances can bias the inferred tidal deformabilities. These results establish tidal resonances as a measurable route for asteroseismology with future detectors.
\end{abstract}

\preprint{INT-PUB-26-022}

\maketitle


\section{Introduction}

Neutron stars harbour the densest known matter in the Universe and serve as nature's laboratories for the elusive equation of state of dense nuclear matter. Since the underlying theory of quantum chromodynamics is intractable in this regime, astrophysical observables are essential for probing the behaviour of matter at supranuclear densities~\cite{Chatziioannou:2024jsr}. Over the last decade, gravitational-wave astronomy has begun to illuminate neutron-star interiors~\cite{LIGOScientific:2025slb}. The discovery of a binary neutron-star merger, known as GW170817, provided the first constraints on the tidal deformability of neutron stars~\cite{LIGOScientific:2017vwq,LIGOScientific:2018hze,LIGOScientific:2018cki}, complementing electromagnetic mass-radius measurements from pulsars, such as those from NASA's NICER mission~\cite{Riley:2019yda,Miller:2019cac,Riley:2021pdl,Miller:2021qha}.

Constraining the properties of dense matter is a key science goal for the third generation of ground-based gravitational-wave interferometers, the Einstein Telescope~\cite{ET:2025xjr} and Cosmic Explorer~\cite{Evans:2023euw}. Their improved sensitivity will allow neutron-star binaries to be observed for longer durations and with much larger signal-to-noise ratios. These enhanced facilities are expected to observe far more mergers, with conservative estimates predicting hundreds of exceptional quality signals annually~\cite{Gupta:2023lga}. With greater sensitivity at high frequencies, where tidal effects become increasingly important, exploiting these observations will require faithful modelling of the neutron-star physics encoded in the gravitational waveform.

Tidal effects provide the main gravitational-wave signature of neutron-star matter before merger~\cite{Flanagan:2007ix}. Each star in a binary system experiences a differential gravitational force from its companion, which raises a tide and accelerates the orbital decay. During the early inspiral, the binary components are well separated and the tidal driving is slow. The stellar response is effectively instantaneous and the static tidal perturbation is quantified by the tidal deformability $\Lambda$~\cite{Hinderer:2007mb,Binnington:2009bb,Damour:2009vw}. Later in the inspiral, the time-evolving tidal field begins to drive the neutron star's natural vibrational modes. This is known as the dynamical tide and it has been studied in detail, partly due to its intrinsic connection with neutron-star seismology~\cite{1994ApJ...426..688R,Lai:1993di,Flanagan:2006sb,Steinhoff:2016rfi,Poisson:2020vap,Pitre:2023xsr,HegadeKR:2024agt,HegadeKR:2025qwj}.

In Newtonian gravity, the connection to asteroseismology is made explicit by decomposing the tidal perturbation using the stellar oscillation modes. Each mode satisfies the familiar equation of motion of a driven harmonic oscillator,
\begin{equation}
    \ddot{a}_\alpha(t) + \omega_\alpha^2 a_\alpha(t) = \frac{Q_\alpha(t)}{\mathcal{E}_\alpha},
    \label{eq:mode}
\end{equation}
where $a_\alpha$ is the mode amplitude, $\omega_\alpha$ is its frequency, $\mathcal{E}_\alpha$ is a normalisation constant and $Q_\alpha$ is the overlap integral, which quantifies how efficiently the tide couples to the oscillation.

Equation~\eqref{eq:mode} reveals two distinct aspects of the dynamical tide. The first is the gradual response, dominated by the fundamental \textit{f} mode, which is now incorporated in several gravitational-waveform models~\cite{Hinderer:2016eia,Schmidt:2019wrl,Abac:2023ujg}. Studies have shown that modelling the \textit{f} mode is necessary to maximise the extracted matter physics from observations~\cite{Pratten:2019sed,Williams:2022vct}, and that neglecting it may introduce systematic biases in equation-of-state inference~\cite{Pratten:2021pro}.

The second aspect of the dynamical tide is transient. In the approach to merger, the tidal driving sweeps upward and can momentarily match the frequency of a low-frequency mode~\cite{Lai:1993di}. When this occurs, the mode is provoked into resonance, growing rapidly in amplitude at the expense of orbital energy. This results in a sudden change in the orbital trajectory.

It remains an open question whether tidal resonances are measurable in a realistic observational scenario. An early study found that second-generation instruments would require a minimum phase shift of $\Delta \Phi \sim 2$ for detection~\cite{Balachandran:2007tu}, larger than what is expected from typical neutron-star oscillations~\cite{Lai:1993di,Flanagan:2006sb,Yu:2016ltf}. However, this conclusion was based on the Fisher information matrix, whose limitations are well known~\cite{Vallisneri:2007ev}. The detectability question has been revisited in recent years. \citet{Ho:2023shr} argued that a serendipitous GW170817-like event observed with Cosmic Explorer could reveal \textit{g}- and \textit{r}-mode resonances, granting access to information about matter composition and stellar spin. In a complementary study, \citet{Revilla-Pena:2025ztm} calculated that a resonance-induced time advance of $O(\qty{1}{\milli\second})$ may be detectable even with second-generation interferometers. While these results are promising, they still rely on approximate sensitivity estimates.

A fully Bayesian study of the detectability of tidal resonances has therefore been lacking. This is what we present for the first time in this Letter. We analyse an astrophysical population of binary neutron-star signals observed by the Einstein Telescope and find that resonantly excited modes can be distinguished. This establishes tidal resonances as an observational target for third-generation interferometers.

\section{Passage through resonance}

Let us set the stage by examining the behaviour of a low-frequency mode as it becomes resonantly excited by the tide. Consider a neutron-star binary on a quasi-circular orbit with angular velocity $\Omega$ inspiralling gradually due to the emission of gravitational radiation. We have a primary of mass $m_1$ and radius $R$ immersed in a tidal field sourced by its companion of mass $m_2$. The binary possesses a chirp mass $\mathcal{M} = (m_1 m_2)^{3 / 5} / (m_1 + m_2)^{1 / 5}$ and a mass ratio $q = m_2 / m_1$. We model the tidally deformed star as a collection of forced harmonic oscillators that satisfy Eq.~\eqref{eq:mode}.

For a non-rotating star in Newtonian gravity, tidal resonance occurs when $|m| \Omega \approx \omega_\alpha$, where $(l, m)$ are the spherical-harmonic quantum numbers. Away from resonance, we neglect the contribution of the oscillation to the tidal response. This is motivated by its small overlap integral, which suppresses the adiabatic response~\cite{1994ApJ...426..688R,Lai:1993di}. During resonance, orbital energy is deposited into the stellar oscillation, perturbing the trajectory of the binary from an orbital phase of $\phi$ to $\phi - \Delta \phi_\alpha$. Under the same approximations, a quadrupolar perturbation sources~\cite{Lai:1993di}
\begin{equation}
    \begin{split}
        \Delta \phi_\alpha &\approx \frac{5 \pi^2}{2048} \left( \frac{c^2 R}{G m_1} \right)^5 \frac{2 q}{1 + q} \left( \frac{\mathcal{I}_{\alpha 2, \pm 2}}{m_1 R^2} \right)^2 \frac{G m_1 / R^3}{\omega_\alpha^2} \\
        &\approx \num{0.04} \, \left( \frac{1.4 M_\odot}{m_1} \right)^4 \left( \frac{R}{\qty{12}{\kilo\metre}} \right)^2 \frac{2 q}{1 + q} \\
        &\quad\times \left[ \frac{\mathcal{I}_{\alpha 2, \pm 2} / (m_1 R^2)}{\num{e-3}} \right]^2 \left[ \frac{\qty{100}{\hertz}}{\omega_\alpha / (2 \pi)} \right]^2,
    \end{split}
    \label{eq:orbital-phase}
\end{equation}
where $G$ and $c$ are Newton's constant and the speed of light, respectively, $I_{\alpha l m}$ is the mode's mass-multipole moment and $\mathcal{I}_{\alpha l m} := I_{\alpha l m} / \sqrt{\mathcal{E}_\alpha / (m_1 R^2)}$. The corresponding gravitational-wave phase  shift is $\Delta \Phi_\alpha = 2 \Delta \phi_\alpha$. When $\Delta \phi_\alpha > 0$, the inspiral is accelerated and the binary merges sooner. This behaviour is generic across mode resonances, with the notable exception of the gravito-magnetic tidal driving of an inertial \textit{r} mode, which instead siphons energy from the spin of the star and injects some into the orbit, producing $\Delta \phi_\alpha < 0$ and delaying the inspiral~\cite{Flanagan:2006sb,Poisson:2020eki}. It is the goal of this Letter to establish whether this small correction is within reach of the Einstein Telescope.

We idealise the resonance as instantaneous. This is justified by comparing the radiation-reaction timescale $\tau_\text{rr} := \Omega / \dot{\Omega}$ to the resonance-crossing timescale $\tau_\res := \sqrt{\pi \tau_{rr} / \Omega}$ through the dimensionless parameter
\begin{equation}
    \epsilon := \frac{\tau_\res}{\tau_\text{rr}} \approx 0.07 \left( \frac{\mathcal{M}}{1.2 M_\odot} \right)^{5 / 6} \left[ \frac{\Omega / (2 \pi)}{\qty{100}{\hertz}} \right]^{5 / 6}.
    \label{eq:timescales}
\end{equation}
Thus, $\epsilon \ll 1$ for orbital frequencies below $\sim \qty{150}{\hertz}$ and the resonance is traversed instantly. Tidal resonances encountered later in the inspiral require a more detailed treatment.

In the stationary-phase approximation, the abrupt change in the orbital motion modifies the gravitational-wave phase by~\cite{Flanagan:2006sb}
\begin{equation}
    \Psi(f) = \Psi_0(f) + \Psi_\alpha(f),
    \label{eq:gravitational-wave-phase}
\end{equation}
where $\Psi_0$ is the phase of a baseline model and
\begin{equation}
    \Psi_\alpha(f) := \Theta(f - f_\alpha) \left( 1 - \frac{f}{f_\alpha} \right) \Delta \Phi_\alpha
    \label{eq:model}
\end{equation}
is the contribution from the tidal resonance. Here, $f$ is the gravitational-wave frequency, $f_\alpha$ is the frequency at resonance and $\Theta$ is the Heaviside step function. In the general case, where multiple oscillation modes are driven into resonance during the inspiral, their phase contributions add linearly in Eq.~\eqref{eq:gravitational-wave-phase}. In the Newtonian approximation, $f_\alpha$ is identically the mode frequency for the quadrupolar resonances considered here. In full general relativity, the effects of redshift and frame-dragging must be taken into account~\cite{Hinderer:2016eia,Steinhoff:2016rfi}.

For the baseline waveform model, we use the approximant \texttt{IMRPhenomXAS\_NRTidalv3}~\cite{Pratten:2020fqn,Abac:2023ujg} as implemented in {\footnotesize LALSuite}~\cite{lalsuite}, which includes the adiabatic \textit{f}-mode contribution to the dynamical tide. Since binary neutron stars are not expected to have appreciable spins~\cite{OShaughnessy:2006uzj}, we neglect precession.

While neutron stars host a notably broad spectrum of oscillation modes~\cite{1988ApJ...325..725M}, we will focus our attention on the simplest scenario where each star in the binary experiences a single resonance. Rather than associating these resonances with a particular mode, we leave their properties parametrised by the gravitational-wave frequencies and phase shifts at resonance, which we denote by $(f_1, \Delta \Phi_1)$ and $(f_2, \Delta \Phi_2)$. In the present Letter, we restrict our focus to positive phase shifts. With the modelling in hand, we now move on to the problem of detection.

\section{Gravitational-wave simulations}

We simulate an astrophysical population of binary neutron-star coalescences observed by the Einstein Telescope over one year. The sources have component masses $m_1, m_2$ that are drawn uniformly from the interval $[1.1, 2.2] M_\odot$ and the tidal deformabilities $\Lambda_1, \Lambda_2$ are determined from the equation of state, which we take to be the model inferred using the collection of datasets denoted ``Set~A'' in Ref.~\cite{Koehn:2024set}. We draw spin magnitudes uniformly on $[0, 0.05]$ with isotropic orientations and retain only the aligned components. The binaries are distributed in redshift $z$ according to
\begin{equation}
    p(z) \propto \frac{\mathcal{R}(z)}{1 + z} \frac{d V_\text{c}}{d z},
\end{equation}
where $V_\text{c}$ is the comoving volume and $\mathcal{R}$ is the merger-rate density. We use the fiducial model of Ref.~\cite{Iorio:2022sgz} with a common-envelope efficiency of $\alpha_\text{CE} = 3$ and assume a flat $\Lambda$CDM cosmology with the parameters inferred from the Planck~2018 data~\cite{Planck:2018vyg}. The binaries' sky positions, inclinations and polarisations are uniform on the sphere. For the mode resonances, we draw the frequencies $f_1, f_2$ uniformly on $[5, 300] \, \unit{\hertz}$ and the phase shifts $\Delta \Phi_1, \Delta \Phi_2$ from a log-uniform distribution on $[10^{-3}, 1]$. These ranges are motivated by Eqs.~\eqref{eq:orbital-phase} and \eqref{eq:timescales}.

We rank the simulated binaries according to optimal signal-to-noise ratio $\rho$ and retain the 200 loudest systems. From our simulations, the resulting signals have $\rho \sim \numrange{60}{450}$. We separate the sources into two populations: one without tidal resonances and another with tidal resonances. The populations have 200 binaries each and differ only in the presence of resonances. This isolates the effect from differences in the astrophysical population and detector noise.

The signals are injected into the triangular Einstein Telescope configuration and coherently added to simulated stationary Gaussian noise following the ET-D design sensitivity~\cite{Hild:2010id}. We analyse the collected data $d(t)$ from a minimum frequency of \qty{5}{\hertz} using the nested-sampling algorithm \textsc{dynesty} with 1000 live points~\cite{Speagle:2019ivv} as implemented in the {\footnotesize Bilby} gravitational-wave inference library~\cite{Ashton:2018jfp}. To speed up the likelihood evaluation, we use relative binning~\cite{Zackay:2018qdy,Narola:2023men}. We validate our results by comparing the relative-binning likelihoods against exact likelihood evaluations and demand that the standard deviation of the log-likelihood error be below $0.1$~\cite{Leslie:2021ssu}.

Since the signals exist in the Einstein Telescope's sensitivity band for $O(\qty{1}{\hour})$, the rotation of the Earth substantially increases the instrument's baseline, allowing the sky location and merger time to be well determined~\cite{Nitz:2021pbr}. In addition, many of these events may have electromagnetic counterparts (like GW170817~\cite{LIGOScientific:2017vwq}), which would also resolve the position well. For these reasons, and for numerical tractability, we fix the sky location and time of coalescence during sampling. Since $\mathcal{M}$ will also be well constrained, we adopt a tight Gaussian prior. The resonance parameters are recovered with slightly broader ranges than the injections: $f_1, f_2$ uniform on $[5, 350] \, \unit{\hertz}$ and $\Delta \Phi_1, \Delta \Phi_2$ log-uniform on $[10^{-4}, 10]$.

\section{Detectability of tidal resonances}

With our simulated injections, we now ask whether the strain data $d$ can distinguish tidal resonances. We consider two competing hypotheses. The first hypothesis $H_\res$ states that $d$ result from a gravitational-wave signal from a merging neutron-star binary where mode resonances took place. The second hypothesis $H_\nores$ claims that $d$ are explained by a binary neutron star with no resonant mode excitations. The Bayesian evidence for hypothesis $H$ is $p(d | H)$. We compare the two hypotheses with the Bayes factor~\cite{Veitch:2009hd}
\begin{equation}
    \bayesfactor := \frac{p(d | H_\res)}{p(d | H_\nores)}
\end{equation}
and adopt $x := \ln \bayesfactor$ as our detection statistic. Positive values of $x$ favour the presence of tidal resonances.

\begin{figure}[h]
    \includegraphics[width=\columnwidth]{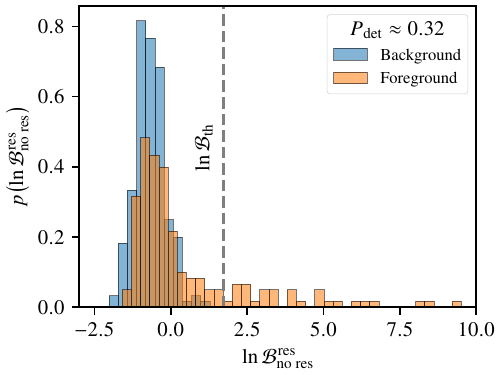}
    \caption{\label{fig:distributions}%
    Distributions of the Bayes factor $\bayesfactor$ comparing the hypothesis that resonantly excited modes are present in the signal against the hypothesis that there are none. The background distribution of sources without tidal resonances (blue histogram) is used to set the detection threshold of $\ln \bayesfactorthreshold \approx 1.73$ (vertical grey line), estimated to correspond to five-sigma significance. The foreground distribution of resonant signals (orange histogram) leads to a detection efficiency of $P_\text{det} \approx 0.32$. Note that the maximum Bayes factor is truncated, although we obtained foreground signals with $\ln \bayesfactor > 10$. The excess of resonant sources above the threshold demonstrates that tidal resonances are detectable in a substantial fraction of the simulated Einstein Telescope observations.}
\end{figure}

To begin with, we must establish the threshold $x_\text{th} = \ln \bayesfactorthreshold$ above which we can claim to have detected mode excitations in the gravitational-wave signal. This is necessary since noise artefacts in the data $d$ can masquerade as resonances and cause the value of $x$ to be elevated. We determine $x_\text{th}$ from a background distribution $p_\text{bg}(x)$, constructed from the 200 simulated signals without tidal resonances. The false-alarm probability is
\begin{equation}
    P_\text{FA}(x_\text{th}) = \int_{x_\text{th}}^\infty p_\text{bg}(x) \, dx,
\end{equation}
representing the chance that a signal with no mode excitations is misclassified as a resonance source. The background (shown in Fig.~\ref{fig:distributions}) is reasonably well behaved, so we may approximate it using a Gaussian kernel density estimator. By demanding $P_\text{FA}(x_\text{th})$ corresponds to five-sigma significance, we estimate $x_\text{th} \approx 1.73$.

Next, we turn our attention to the foreground distribution $p_\text{fg}(x)$ of the 200 sources with resonantly excited modes. From $p_\text{fg}(x)$, we may deduce the detection efficiency
\begin{equation}
    P_\text{det}(x_\text{th}) = \int_{x_\text{th}}^\infty p_\text{fg}(x) \, dx,
\end{equation}
which quantifies the rate at which foreground sources are distinguishable from the background. We find that roughly one in three signals ($P_\text{det} \approx 0.32$) has detectable resonances, demonstrating that resonant modes can be distinguished in a sizeable subset of the loudest Einstein Telescope events. The background and foreground distributions are both shown in Fig.~\ref{fig:distributions}.

\begin{figure}[h]
    \includegraphics[width=\columnwidth]{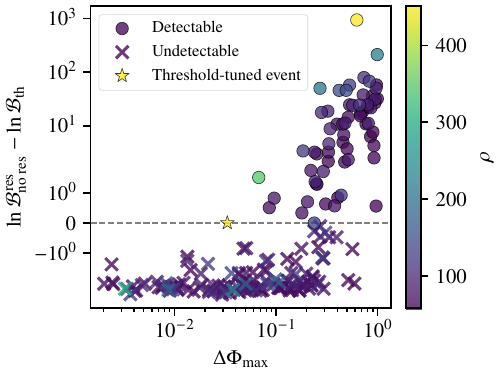}
    \caption{\label{fig:log-bayes-factor}%
    The Bayes factor $\bayesfactor$ as a function of the larger gravitational-wave phase shift $\Delta \Phi_\text{max}$ for the foreground sources. Each event is coloured by its optimal signal-to-noise ratio $\rho$. Events that exceed the threshold $\bayesfactorthreshold$ (horizontal grey line) are classified as detectable (circles), while those below are classified as undetectable (crosses). The detectable sources are primarily concentrated at large $\Delta \Phi_\text{max}$, indicating that the phase shift drives the detectability. We show a tuned source (star) constructed from the loudest signal by reducing its shifts until $\Delta \Phi_\text{max} \approx 0.03$. This provides an estimate of the minimum detectable phase shift with the Einstein Telescope.}
\end{figure}

Figure~\ref{fig:log-bayes-factor} shows the foreground sources, where it can be seen that the phase shift is the dominant contributor to detectability. We analyse the detection efficiency as a function of $\Delta \Phi_\text{max} := \text{max}(\Delta \Phi_1, \Delta \Phi_2)$. At $\Delta \Phi_\text{max} \approx 0.07$, we find that one in five sources ($P_\text{det} \approx 0.2$) is detectable. This doubles to $P_\text{det} \approx 0.4$ once the shifts reach the level of $\Delta \Phi_\text{max} \approx 0.22$. We find no discernible dependence on the resonance frequencies over the sampled range $f_1, f_2 \in [5, 300] \, \unit{\hertz}$.

To estimate the minimum detectable phase shift in a favourable system, we modify the resonance parameters of the loudest signal in our catalogue. Assuming the same noise realisation and source parameters, we reduce its phase shifts until the event approaches the detection threshold. The signal remains detectable down to $\Delta \Phi_\text{max} \approx 0.03$. This event is marked in Fig.~\ref{fig:log-bayes-factor}.

\begin{figure}[h]
    \includegraphics[width=\columnwidth]{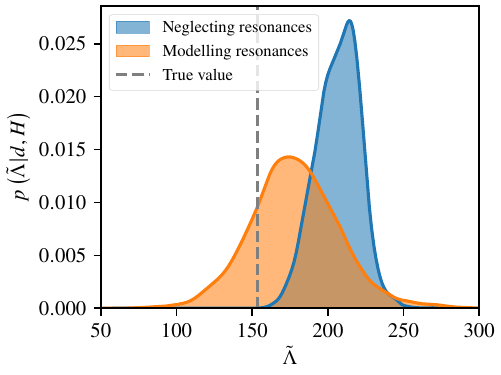}
    \caption{\label{fig:posterior}%
    Posterior probability distribution $p(\tilde{\Lambda} | d, H)$ on the mass-weighted tidal deformability $\tilde{\Lambda}$ for a representative detectable event. The source has a maximum phase shift of $\Delta \Phi_\text{max} \approx 0.24$ and signal-to-noise ratio of $\rho \approx 83$. The true value (vertical grey line) is recovered when resonances are modelled (orange distribution), but not when they are omitted (blue distribution). This example demonstrates that neglecting the presence of mode resonances can bias tidal-deformability measurements for large phase shifts.}
\end{figure}

Lastly, we can use our results to explore the impact of resonance modelling on parameter estimation. We find cases in which neglecting the resonances biases the recovery of the mass-weighted tidal deformability $\tilde{\Lambda}$. This typically occurs for phase shifts of $\Delta \Phi_\text{max} \gtrsim 0.2$, where $\tilde{\Lambda}$ is biased towards larger values, consistent with our focus on positive phase shifts. Conversely, for negative shifts that delay the inspiral, we would expect the bias to be towards smaller $\tilde{\Lambda}$. A representative example is presented in Fig.~\ref{fig:posterior}. Predictably, the bias becomes less pronounced towards lower $\rho$ and smaller resonances. We also find less prominent biases in the recovery of the chirp mass $\mathcal{M}$, while biases occasionally occur for the sub-dominant tidal parameter $\delta \tilde{\Lambda}$, which is comparatively weakly constrained.

\section{Implications for neutron-star modes}

Before we conclude, let us discuss the low-frequency oscillation modes that may achieve these phase shifts. Composition gradients in the neutron-star interior provide the buoyancy that supports \textit{g}-mode oscillations~\cite{1992ApJ...395..240R}. Recent relativistic calculations with realistic nuclear-matter models find that these perturbations have frequencies $\omega_\alpha / (2 \pi) \sim \qty{100}{\hertz}$ and quadrupole moments $\mathcal{I}_{\alpha 2, \pm 2} / (m_1 R^2) \sim \numrange{e-3}{e-5}$, depending crucially on the degree of composition gradients in the equation of state~\cite{Counsell:2024pua}. Using the canonical scaling in Eq.~\eqref{eq:orbital-phase}, this amounts to phase shifts up to $\Delta \Phi_\alpha \sim 0.08$, within the range probed by our study.

Inertial modes provide another plausible class of detectable resonances. They arise due to the stellar rotation~\cite{Lockitch:1998nq}. Recent Newtonian calculations have identified four inertial modes, including the \textit{r} mode, that become resonantly excited during an inspiral~\cite{Poisson:2020eki}. For simple stellar models, phase shifts of $|\Delta \Phi_\alpha| \sim \numrange{0.02}{0.20}$ are within reach, depending on the alignment of the star's spin with the orbital angular momentum. It should be noted that the \textit{r} mode gives rise to $\Delta \Phi_\alpha < 0$, which was not the sign we considered in this Letter.

By contrast, the resonant excitation of a core-crust interface mode, which may produce resonant shattering flares~\cite{Tsang:2011ad}, appears less promising for gravitational-wave detection. A recent relativistic study found that these perturbations produce $\Delta \Phi_\alpha \approx 0.002$~\cite{Gao:2025aqo}, below the sensitivity reached in our study.

There has been some recent discussion of the interface modes (also known as discontinuity \textit{g} modes) due to a sharp phase transition to deconfined quark matter~\cite{Miao:2023jqe,Counsell:2025hcv,Pereira:2025xsi}. These modes have higher frequencies, beyond what we could reliably model with our resonance treatment of Eq.~\eqref{eq:model}. Improvements in the modelling of resonances will be required to study their detectability~\cite{Yu:2024uxt,Pnigouras:2025muo}. Thus, the present results point primarily to composition \textit{g} modes and inertial modes as the most plausible low-frequency resonances for Einstein Telescope asteroseismology.

\section{Conclusions}

We have conducted the first fully Bayesian investigation into the Einstein Telescope's ability to detect resonantly excited oscillation modes in the gravitational-wave signal from binary neutron-star coalescences. With one year of observations, we found that the gravitational-wave observatory will be able to distinguish mode resonances. Our results show that the Einstein Telescope is sensitive to resonant modes with frequencies of $\qtyrange{5}{300}{\hertz}$ and phase shifts down to $\Delta \Phi \approx 0.03$, under favourable conditions. These conclusions are not expected to depend sensitively on the proposed detector geometries.

This sensitivity reaches the range expected for several low-frequency neutron-star oscillation modes. In particular, \textit{g} modes and inertial modes can produce phase shifts comparable to those found to be detectable here. While crust-core interface modes are expected to produce smaller resonances, the addition of Cosmic Explorer to the network may bring these oscillations within observational reach. These results therefore identify resonant tides as a measurable route to neutron-star asteroseismology with third-generation detectors.

We have also demonstrated that resonances can affect parameter inference if they are omitted from the modelling. For sufficiently large phase shifts, the missing resonant phase evolution can be partially absorbed by the mass-weighted tidal deformability, biasing the recovery.

The natural next step is to connect the mode parameters to dense-matter properties and infer these characteristics during parameter estimation. This would establish how measurements can be translated into equation-of-state constraints. \\


\begin{acknowledgments}
    The authors thank Filippo Santoliquido for discussions on the astrophysical population model used in this work.
    F.\,G. acknowledges funding from the European Union's Horizon Europe research and innovation programme under the Marie Sk{\l}odowska-Curie Grant Agreement No.~101151301.
    H.\,N., T.\,W., T.\,H. and C.\,V.\,D.\,B. are supported by the research programme of the Netherlands Organization for Scientific Research through Grant No.~OCENW.XL21.XL21.038.
    P.\,T.\,H.\,P. is supported by the research programme of the Netherlands Organization for Scientific Research under Grant No.~VI.Veni.232.021.
    This work used the Dutch national e-infrastructure with the support of the SURF Cooperative using Grant No.~EINF-15521.
    F.\,G. and P.\,T.\,H.\,P. thank the Institute for Nuclear Theory at the University of Washington for its kind hospitality and stimulating research environment, where part of this work was conducted. This research was supported in part by the Institute's U.S. Department of Energy Grant No.~DE-FG02-00ER41132.

    The data that support the findings of this article are publicly available~\cite{zenodo}.
\end{acknowledgments}


\bibliography{refs}

\begin{thebibliography}{62}%
\makeatletter
\providecommand \@ifxundefined [1]{%
 \@ifx{#1\undefined}
}%
\providecommand \@ifnum [1]{%
 \ifnum #1\expandafter \@firstoftwo
 \else \expandafter \@secondoftwo
 \fi
}%
\providecommand \@ifx [1]{%
 \ifx #1\expandafter \@firstoftwo
 \else \expandafter \@secondoftwo
 \fi
}%
\providecommand \natexlab [1]{#1}%
\providecommand \enquote  [1]{``#1''}%
\providecommand \bibnamefont  [1]{#1}%
\providecommand \bibfnamefont [1]{#1}%
\providecommand \citenamefont [1]{#1}%
\providecommand \href@noop [0]{\@secondoftwo}%
\providecommand \href [0]{\begingroup \@sanitize@url \@href}%
\providecommand \@href[1]{\@@startlink{#1}\@@href}%
\providecommand \@@href[1]{\endgroup#1\@@endlink}%
\providecommand \@sanitize@url [0]{\catcode `\\12\catcode `\$12\catcode
  `\&12\catcode `\#12\catcode `\^12\catcode `\_12\catcode `\%12\relax}%
\providecommand \@@startlink[1]{}%
\providecommand \@@endlink[0]{}%
\providecommand \url  [0]{\begingroup\@sanitize@url \@url }%
\providecommand \@url [1]{\endgroup\@href {#1}{\urlprefix }}%
\providecommand \urlprefix  [0]{URL }%
\providecommand \Eprint [0]{\href }%
\providecommand \doibase [0]{https://doi.org/}%
\providecommand \selectlanguage [0]{\@gobble}%
\providecommand \bibinfo  [0]{\@secondoftwo}%
\providecommand \bibfield  [0]{\@secondoftwo}%
\providecommand \translation [1]{[#1]}%
\providecommand \BibitemOpen [0]{}%
\providecommand \bibitemStop [0]{}%
\providecommand \bibitemNoStop [0]{.\EOS\space}%
\providecommand \EOS [0]{\spacefactor3000\relax}%
\providecommand \BibitemShut  [1]{\csname bibitem#1\endcsname}%
\let\auto@bib@innerbib\@empty
\bibitem [{\citenamefont {Chatziioannou}\ \emph {et~al.}(2025)\citenamefont
  {Chatziioannou}, \citenamefont {Cromartie}, \citenamefont {Gandolfi},
  \citenamefont {Tews}, \citenamefont {Radice}, \citenamefont {Steiner},\ and\
  \citenamefont {Watts}}]{Chatziioannou:2024jsr}%
  \BibitemOpen
  \bibfield  {author} {\bibinfo {author} {\bibfnamefont {K.}~\bibnamefont
  {Chatziioannou}}, \bibinfo {author} {\bibfnamefont {H.~T.}\ \bibnamefont
  {Cromartie}}, \bibinfo {author} {\bibfnamefont {S.}~\bibnamefont {Gandolfi}},
  \bibinfo {author} {\bibfnamefont {I.}~\bibnamefont {Tews}}, \bibinfo {author}
  {\bibfnamefont {D.}~\bibnamefont {Radice}}, \bibinfo {author} {\bibfnamefont
  {A.~W.}\ \bibnamefont {Steiner}},\ and\ \bibinfo {author} {\bibfnamefont
  {A.~L.}\ \bibnamefont {Watts}},\ }\href {https://doi.org/10.1103/ymsq-cfcw}
  {\bibfield  {journal} {\bibinfo  {journal} {\rmp}\ }\textbf {\bibinfo
  {volume} {97}},\ \bibinfo {pages} {045007} (\bibinfo {year}
  {2025})}\BibitemShut {NoStop}%
\bibitem [{\citenamefont {Abac}\ \emph
  {et~al.}(2026{\natexlab{a}})\citenamefont {Abac} \emph
  {et~al.}}]{LIGOScientific:2025slb}%
  \BibitemOpen
  \bibfield  {author} {\bibinfo {author} {\bibfnamefont {A.~G.}\ \bibnamefont
  {Abac}} \emph {et~al.} (\bibinfo {collaboration} {LIGO Scientific, VIRGO,
  KAGRA}),\ }\href {https://doi.org/10.3847/2041-8213/ae2c74} {\bibfield
  {journal} {\bibinfo  {journal} {\apj}\ }\textbf {\bibinfo {volume} {1004}},\
  \bibinfo {pages} {L22} (\bibinfo {year} {2026}{\natexlab{a}})}\BibitemShut
  {NoStop}%
\bibitem [{\citenamefont {Abbott}\ \emph {et~al.}(2017)\citenamefont {Abbott}
  \emph {et~al.}}]{LIGOScientific:2017vwq}%
  \BibitemOpen
  \bibfield  {author} {\bibinfo {author} {\bibfnamefont {B.~P.}\ \bibnamefont
  {Abbott}} \emph {et~al.} (\bibinfo {collaboration} {LIGO Scientific,
  Virgo}),\ }\href {https://doi.org/10.1103/PhysRevLett.119.161101} {\bibfield
  {journal} {\bibinfo  {journal} {\prl}\ }\textbf {\bibinfo {volume} {119}},\
  \bibinfo {pages} {161101} (\bibinfo {year} {2017})}\BibitemShut {NoStop}%
\bibitem [{\citenamefont {Abbott}\ \emph {et~al.}(2019)\citenamefont {Abbott}
  \emph {et~al.}}]{LIGOScientific:2018hze}%
  \BibitemOpen
  \bibfield  {author} {\bibinfo {author} {\bibfnamefont {B.~P.}\ \bibnamefont
  {Abbott}} \emph {et~al.} (\bibinfo {collaboration} {LIGO Scientific,
  Virgo}),\ }\href {https://doi.org/10.1103/PhysRevX.9.011001} {\bibfield
  {journal} {\bibinfo  {journal} {\prx}\ }\textbf {\bibinfo {volume} {9}},\
  \bibinfo {pages} {011001} (\bibinfo {year} {2019})}\BibitemShut {NoStop}%
\bibitem [{\citenamefont {Abbott}\ \emph {et~al.}(2018)\citenamefont {Abbott}
  \emph {et~al.}}]{LIGOScientific:2018cki}%
  \BibitemOpen
  \bibfield  {author} {\bibinfo {author} {\bibfnamefont {B.~P.}\ \bibnamefont
  {Abbott}} \emph {et~al.} (\bibinfo {collaboration} {LIGO Scientific,
  Virgo}),\ }\href {https://doi.org/10.1103/PhysRevLett.121.161101} {\bibfield
  {journal} {\bibinfo  {journal} {\prl}\ }\textbf {\bibinfo {volume} {121}},\
  \bibinfo {pages} {161101} (\bibinfo {year} {2018})}\BibitemShut {NoStop}%
\bibitem [{\citenamefont {Riley}\ \emph {et~al.}(2019)\citenamefont {Riley}
  \emph {et~al.}}]{Riley:2019yda}%
  \BibitemOpen
  \bibfield  {author} {\bibinfo {author} {\bibfnamefont {T.~E.}\ \bibnamefont
  {Riley}} \emph {et~al.},\ }\href {https://doi.org/10.3847/2041-8213/ab481c}
  {\bibfield  {journal} {\bibinfo  {journal} {\apj}\ }\textbf {\bibinfo
  {volume} {887}},\ \bibinfo {pages} {L21} (\bibinfo {year}
  {2019})}\BibitemShut {NoStop}%
\bibitem [{\citenamefont {Miller}\ \emph {et~al.}(2019)\citenamefont {Miller}
  \emph {et~al.}}]{Miller:2019cac}%
  \BibitemOpen
  \bibfield  {author} {\bibinfo {author} {\bibfnamefont {M.~C.}\ \bibnamefont
  {Miller}} \emph {et~al.},\ }\href {https://doi.org/10.3847/2041-8213/ab50c5}
  {\bibfield  {journal} {\bibinfo  {journal} {\apj}\ }\textbf {\bibinfo
  {volume} {887}},\ \bibinfo {pages} {L24} (\bibinfo {year}
  {2019})}\BibitemShut {NoStop}%
\bibitem [{\citenamefont {Riley}\ \emph {et~al.}(2021)\citenamefont {Riley}
  \emph {et~al.}}]{Riley:2021pdl}%
  \BibitemOpen
  \bibfield  {author} {\bibinfo {author} {\bibfnamefont {T.~E.}\ \bibnamefont
  {Riley}} \emph {et~al.},\ }\href {https://doi.org/10.3847/2041-8213/ac0a81}
  {\bibfield  {journal} {\bibinfo  {journal} {\apj}\ }\textbf {\bibinfo
  {volume} {918}},\ \bibinfo {pages} {L27} (\bibinfo {year}
  {2021})}\BibitemShut {NoStop}%
\bibitem [{\citenamefont {Miller}\ \emph {et~al.}(2021)\citenamefont {Miller}
  \emph {et~al.}}]{Miller:2021qha}%
  \BibitemOpen
  \bibfield  {author} {\bibinfo {author} {\bibfnamefont {M.~C.}\ \bibnamefont
  {Miller}} \emph {et~al.},\ }\href {https://doi.org/10.3847/2041-8213/ac089b}
  {\bibfield  {journal} {\bibinfo  {journal} {\apj}\ }\textbf {\bibinfo
  {volume} {918}},\ \bibinfo {pages} {L28} (\bibinfo {year}
  {2021})}\BibitemShut {NoStop}%
\bibitem [{\citenamefont {Abac}\ \emph
  {et~al.}(2026{\natexlab{b}})\citenamefont {Abac} \emph
  {et~al.}}]{ET:2025xjr}%
  \BibitemOpen
  \bibfield  {author} {\bibinfo {author} {\bibfnamefont {A.}~\bibnamefont
  {Abac}} \emph {et~al.} (\bibinfo {collaboration} {ET}),\ }\href
  {https://doi.org/10.1088/1475-7516/2026/03/081} {\bibfield  {journal}
  {\bibinfo  {journal} {\jcap}\ }\textbf {\bibinfo {volume} {03}},\ \bibinfo
  {pages} {081} (\bibinfo {year} {2026}{\natexlab{b}})}\BibitemShut {NoStop}%
\bibitem [{\citenamefont {Evans}\ \emph {et~al.}(2023)\citenamefont {Evans}
  \emph {et~al.}}]{Evans:2023euw}%
  \BibitemOpen
  \bibfield  {author} {\bibinfo {author} {\bibfnamefont {M.}~\bibnamefont
  {Evans}} \emph {et~al.},\ }\Eprint {https://arxiv.org/abs/2306.13745}
  {arXiv:2306.13745 [astro-ph.IM]}  (\bibinfo {year} {2023})\BibitemShut
  {NoStop}%
\bibitem [{\citenamefont {Gupta}\ \emph {et~al.}(2024)\citenamefont {Gupta}
  \emph {et~al.}}]{Gupta:2023lga}%
  \BibitemOpen
  \bibfield  {author} {\bibinfo {author} {\bibfnamefont {I.}~\bibnamefont
  {Gupta}} \emph {et~al.},\ }\href {https://doi.org/10.1088/1361-6382/ad7b99}
  {\bibfield  {journal} {\bibinfo  {journal} {\cqg}\ }\textbf {\bibinfo
  {volume} {41}},\ \bibinfo {pages} {245001} (\bibinfo {year}
  {2024})}\BibitemShut {NoStop}%
\bibitem [{\citenamefont {Flanagan}\ and\ \citenamefont
  {Hinderer}(2008)}]{Flanagan:2007ix}%
  \BibitemOpen
  \bibfield  {author} {\bibinfo {author} {\bibfnamefont {E.~E.}\ \bibnamefont
  {Flanagan}}\ and\ \bibinfo {author} {\bibfnamefont {T.}~\bibnamefont
  {Hinderer}},\ }\href {https://doi.org/10.1103/PhysRevD.77.021502} {\bibfield
  {journal} {\bibinfo  {journal} {\prd}\ }\textbf {\bibinfo {volume} {77}},\
  \bibinfo {pages} {021502} (\bibinfo {year} {2008})}\BibitemShut {NoStop}%
\bibitem [{\citenamefont {Hinderer}(2008)}]{Hinderer:2007mb}%
  \BibitemOpen
  \bibfield  {author} {\bibinfo {author} {\bibfnamefont {T.}~\bibnamefont
  {Hinderer}},\ }\href {https://doi.org/10.1086/533487} {\bibfield  {journal}
  {\bibinfo  {journal} {\apj}\ }\textbf {\bibinfo {volume} {677}},\ \bibinfo
  {pages} {1216} (\bibinfo {year} {2008})},\ \bibinfo {note} {[Erratum: \apj\
  \textbf{697}, 964 (2009)]}\BibitemShut {NoStop}%
\bibitem [{\citenamefont {Binnington}\ and\ \citenamefont
  {Poisson}(2009)}]{Binnington:2009bb}%
  \BibitemOpen
  \bibfield  {author} {\bibinfo {author} {\bibfnamefont {T.}~\bibnamefont
  {Binnington}}\ and\ \bibinfo {author} {\bibfnamefont {E.}~\bibnamefont
  {Poisson}},\ }\href {https://doi.org/10.1103/PhysRevD.80.084018} {\bibfield
  {journal} {\bibinfo  {journal} {\prd}\ }\textbf {\bibinfo {volume} {80}},\
  \bibinfo {pages} {084018} (\bibinfo {year} {2009})}\BibitemShut {NoStop}%
\bibitem [{\citenamefont {Damour}\ and\ \citenamefont
  {Nagar}(2009)}]{Damour:2009vw}%
  \BibitemOpen
  \bibfield  {author} {\bibinfo {author} {\bibfnamefont {T.}~\bibnamefont
  {Damour}}\ and\ \bibinfo {author} {\bibfnamefont {A.}~\bibnamefont {Nagar}},\
  }\href {https://doi.org/10.1103/PhysRevD.80.084035} {\bibfield  {journal}
  {\bibinfo  {journal} {\prd}\ }\textbf {\bibinfo {volume} {80}},\ \bibinfo
  {pages} {084035} (\bibinfo {year} {2009})}\BibitemShut {NoStop}%
\bibitem [{\citenamefont {{Reisenegger}}\ and\ \citenamefont
  {{Goldreich}}(1994)}]{1994ApJ...426..688R}%
  \BibitemOpen
  \bibfield  {author} {\bibinfo {author} {\bibfnamefont {A.}~\bibnamefont
  {{Reisenegger}}}\ and\ \bibinfo {author} {\bibfnamefont {P.}~\bibnamefont
  {{Goldreich}}},\ }\href {https://doi.org/10.1086/174105} {\bibfield
  {journal} {\bibinfo  {journal} {\apj}\ }\textbf {\bibinfo {volume} {426}},\
  \bibinfo {pages} {688} (\bibinfo {year} {1994})}\BibitemShut {NoStop}%
\bibitem [{\citenamefont {Lai}(1994)}]{Lai:1993di}%
  \BibitemOpen
  \bibfield  {author} {\bibinfo {author} {\bibfnamefont {D.}~\bibnamefont
  {Lai}},\ }\href {https://doi.org/10.1093/mnras/270.3.611} {\bibfield
  {journal} {\bibinfo  {journal} {\mnras}\ }\textbf {\bibinfo {volume} {270}},\
  \bibinfo {pages} {611} (\bibinfo {year} {1994})}\BibitemShut {NoStop}%
\bibitem [{\citenamefont {Flanagan}\ and\ \citenamefont
  {Racine}(2007)}]{Flanagan:2006sb}%
  \BibitemOpen
  \bibfield  {author} {\bibinfo {author} {\bibfnamefont {E.~E.}\ \bibnamefont
  {Flanagan}}\ and\ \bibinfo {author} {\bibfnamefont {E.}~\bibnamefont
  {Racine}},\ }\href {https://doi.org/10.1103/PhysRevD.75.044001} {\bibfield
  {journal} {\bibinfo  {journal} {\prd}\ }\textbf {\bibinfo {volume} {75}},\
  \bibinfo {pages} {044001} (\bibinfo {year} {2007})}\BibitemShut {NoStop}%
\bibitem [{\citenamefont {Steinhoff}\ \emph {et~al.}(2016)\citenamefont
  {Steinhoff}, \citenamefont {Hinderer}, \citenamefont {Buonanno},\ and\
  \citenamefont {Taracchini}}]{Steinhoff:2016rfi}%
  \BibitemOpen
  \bibfield  {author} {\bibinfo {author} {\bibfnamefont {J.}~\bibnamefont
  {Steinhoff}}, \bibinfo {author} {\bibfnamefont {T.}~\bibnamefont {Hinderer}},
  \bibinfo {author} {\bibfnamefont {A.}~\bibnamefont {Buonanno}},\ and\
  \bibinfo {author} {\bibfnamefont {A.}~\bibnamefont {Taracchini}},\ }\href
  {https://doi.org/10.1103/PhysRevD.94.104028} {\bibfield  {journal} {\bibinfo
  {journal} {\prd}\ }\textbf {\bibinfo {volume} {94}},\ \bibinfo {pages}
  {104028} (\bibinfo {year} {2016})}\BibitemShut {NoStop}%
\bibitem [{\citenamefont {Poisson}(2021)}]{Poisson:2020vap}%
  \BibitemOpen
  \bibfield  {author} {\bibinfo {author} {\bibfnamefont {E.}~\bibnamefont
  {Poisson}},\ }\href {https://doi.org/10.1103/PhysRevD.103.064023} {\bibfield
  {journal} {\bibinfo  {journal} {\prd}\ }\textbf {\bibinfo {volume} {103}},\
  \bibinfo {pages} {064023} (\bibinfo {year} {2021})}\BibitemShut {NoStop}%
\bibitem [{\citenamefont {Pitre}\ and\ \citenamefont
  {Poisson}(2024)}]{Pitre:2023xsr}%
  \BibitemOpen
  \bibfield  {author} {\bibinfo {author} {\bibfnamefont {T.}~\bibnamefont
  {Pitre}}\ and\ \bibinfo {author} {\bibfnamefont {E.}~\bibnamefont
  {Poisson}},\ }\href {https://doi.org/10.1103/PhysRevD.109.064004} {\bibfield
  {journal} {\bibinfo  {journal} {\prd}\ }\textbf {\bibinfo {volume} {109}},\
  \bibinfo {pages} {064004} (\bibinfo {year} {2024})}\BibitemShut {NoStop}%
\bibitem [{\citenamefont {Hegade K.~R.}\ \emph {et~al.}(2024)\citenamefont
  {Hegade K.~R.}, \citenamefont {Ripley},\ and\ \citenamefont
  {Yunes}}]{HegadeKR:2024agt}%
  \BibitemOpen
  \bibfield  {author} {\bibinfo {author} {\bibfnamefont {A.}~\bibnamefont
  {Hegade K.~R.}}, \bibinfo {author} {\bibfnamefont {J.~L.}\ \bibnamefont
  {Ripley}},\ and\ \bibinfo {author} {\bibfnamefont {N.}~\bibnamefont
  {Yunes}},\ }\href {https://doi.org/10.1103/PhysRevD.109.104064} {\bibfield
  {journal} {\bibinfo  {journal} {\prd}\ }\textbf {\bibinfo {volume} {109}},\
  \bibinfo {pages} {104064} (\bibinfo {year} {2024})}\BibitemShut {NoStop}%
\bibitem [{\citenamefont {Hegade K.~R.}\ \emph {et~al.}(2026)\citenamefont
  {Hegade K.~R.}, \citenamefont {Kwon}, \citenamefont {Venumadhav},
  \citenamefont {Yu},\ and\ \citenamefont {Yunes}}]{HegadeKR:2025qwj}%
  \BibitemOpen
  \bibfield  {author} {\bibinfo {author} {\bibfnamefont {A.}~\bibnamefont
  {Hegade K.~R.}}, \bibinfo {author} {\bibfnamefont {K.~J.}\ \bibnamefont
  {Kwon}}, \bibinfo {author} {\bibfnamefont {T.}~\bibnamefont {Venumadhav}},
  \bibinfo {author} {\bibfnamefont {H.}~\bibnamefont {Yu}},\ and\ \bibinfo
  {author} {\bibfnamefont {N.}~\bibnamefont {Yunes}},\ }\href
  {https://doi.org/10.1103/1wdp-6x27} {\bibfield  {journal} {\bibinfo
  {journal} {\prl}\ }\textbf {\bibinfo {volume} {136}},\ \bibinfo {pages}
  {071401} (\bibinfo {year} {2026})}\BibitemShut {NoStop}%
\bibitem [{\citenamefont {Hinderer}\ \emph {et~al.}(2016)\citenamefont
  {Hinderer} \emph {et~al.}}]{Hinderer:2016eia}%
  \BibitemOpen
  \bibfield  {author} {\bibinfo {author} {\bibfnamefont {T.}~\bibnamefont
  {Hinderer}} \emph {et~al.},\ }\href
  {https://doi.org/10.1103/PhysRevLett.116.181101} {\bibfield  {journal}
  {\bibinfo  {journal} {\prl}\ }\textbf {\bibinfo {volume} {116}},\ \bibinfo
  {pages} {181101} (\bibinfo {year} {2016})}\BibitemShut {NoStop}%
\bibitem [{\citenamefont {Schmidt}\ and\ \citenamefont
  {Hinderer}(2019)}]{Schmidt:2019wrl}%
  \BibitemOpen
  \bibfield  {author} {\bibinfo {author} {\bibfnamefont {P.}~\bibnamefont
  {Schmidt}}\ and\ \bibinfo {author} {\bibfnamefont {T.}~\bibnamefont
  {Hinderer}},\ }\href {https://doi.org/10.1103/PhysRevD.100.021501} {\bibfield
   {journal} {\bibinfo  {journal} {\prd}\ }\textbf {\bibinfo {volume} {100}},\
  \bibinfo {pages} {021501} (\bibinfo {year} {2019})}\BibitemShut {NoStop}%
\bibitem [{\citenamefont {Abac}\ \emph {et~al.}(2024)\citenamefont {Abac},
  \citenamefont {Dietrich}, \citenamefont {Buonanno}, \citenamefont
  {Steinhoff},\ and\ \citenamefont {Ujevic}}]{Abac:2023ujg}%
  \BibitemOpen
  \bibfield  {author} {\bibinfo {author} {\bibfnamefont {A.}~\bibnamefont
  {Abac}}, \bibinfo {author} {\bibfnamefont {T.}~\bibnamefont {Dietrich}},
  \bibinfo {author} {\bibfnamefont {A.}~\bibnamefont {Buonanno}}, \bibinfo
  {author} {\bibfnamefont {J.}~\bibnamefont {Steinhoff}},\ and\ \bibinfo
  {author} {\bibfnamefont {M.}~\bibnamefont {Ujevic}},\ }\href
  {https://doi.org/10.1103/PhysRevD.109.024062} {\bibfield  {journal} {\bibinfo
   {journal} {\prd}\ }\textbf {\bibinfo {volume} {109}},\ \bibinfo {pages}
  {024062} (\bibinfo {year} {2024})}\BibitemShut {NoStop}%
\bibitem [{\citenamefont {Pratten}\ \emph
  {et~al.}(2020{\natexlab{a}})\citenamefont {Pratten}, \citenamefont
  {Schmidt},\ and\ \citenamefont {Hinderer}}]{Pratten:2019sed}%
  \BibitemOpen
  \bibfield  {author} {\bibinfo {author} {\bibfnamefont {G.}~\bibnamefont
  {Pratten}}, \bibinfo {author} {\bibfnamefont {P.}~\bibnamefont {Schmidt}},\
  and\ \bibinfo {author} {\bibfnamefont {T.}~\bibnamefont {Hinderer}},\ }\href
  {https://doi.org/10.1038/s41467-020-15984-5} {\bibfield  {journal} {\bibinfo
  {journal} {\natcom}\ }\textbf {\bibinfo {volume} {11}},\ \bibinfo {pages}
  {2553} (\bibinfo {year} {2020}{\natexlab{a}})}\BibitemShut {NoStop}%
\bibitem [{\citenamefont {Williams}\ \emph {et~al.}(2022)\citenamefont
  {Williams}, \citenamefont {Pratten},\ and\ \citenamefont
  {Schmidt}}]{Williams:2022vct}%
  \BibitemOpen
  \bibfield  {author} {\bibinfo {author} {\bibfnamefont {N.}~\bibnamefont
  {Williams}}, \bibinfo {author} {\bibfnamefont {G.}~\bibnamefont {Pratten}},\
  and\ \bibinfo {author} {\bibfnamefont {P.}~\bibnamefont {Schmidt}},\ }\href
  {https://doi.org/10.1103/PhysRevD.105.123032} {\bibfield  {journal} {\bibinfo
   {journal} {\prd}\ }\textbf {\bibinfo {volume} {105}},\ \bibinfo {pages}
  {123032} (\bibinfo {year} {2022})}\BibitemShut {NoStop}%
\bibitem [{\citenamefont {Pratten}\ \emph {et~al.}(2022)\citenamefont
  {Pratten}, \citenamefont {Schmidt},\ and\ \citenamefont
  {Williams}}]{Pratten:2021pro}%
  \BibitemOpen
  \bibfield  {author} {\bibinfo {author} {\bibfnamefont {G.}~\bibnamefont
  {Pratten}}, \bibinfo {author} {\bibfnamefont {P.}~\bibnamefont {Schmidt}},\
  and\ \bibinfo {author} {\bibfnamefont {N.}~\bibnamefont {Williams}},\ }\href
  {https://doi.org/10.1103/PhysRevLett.129.081102} {\bibfield  {journal}
  {\bibinfo  {journal} {\prl}\ }\textbf {\bibinfo {volume} {129}},\ \bibinfo
  {pages} {081102} (\bibinfo {year} {2022})}\BibitemShut {NoStop}%
\bibitem [{\citenamefont {Balachandran}\ and\ \citenamefont
  {Flanagan}(2007)}]{Balachandran:2007tu}%
  \BibitemOpen
  \bibfield  {author} {\bibinfo {author} {\bibfnamefont {P.}~\bibnamefont
  {Balachandran}}\ and\ \bibinfo {author} {\bibfnamefont {E.~E.}\ \bibnamefont
  {Flanagan}},\ }\Eprint {https://arxiv.org/abs/gr-qc/0701076}
  {arXiv:gr-qc/0701076}  (\bibinfo {year} {2007})\BibitemShut {NoStop}%
\bibitem [{\citenamefont {Yu}\ and\ \citenamefont
  {Weinberg}(2017)}]{Yu:2016ltf}%
  \BibitemOpen
  \bibfield  {author} {\bibinfo {author} {\bibfnamefont {H.}~\bibnamefont
  {Yu}}\ and\ \bibinfo {author} {\bibfnamefont {N.~N.}\ \bibnamefont
  {Weinberg}},\ }\href {https://doi.org/10.1093/mnras/stw2552} {\bibfield
  {journal} {\bibinfo  {journal} {\mnras}\ }\textbf {\bibinfo {volume} {464}},\
  \bibinfo {pages} {2622} (\bibinfo {year} {2017})}\BibitemShut {NoStop}%
\bibitem [{\citenamefont {Vallisneri}(2008)}]{Vallisneri:2007ev}%
  \BibitemOpen
  \bibfield  {author} {\bibinfo {author} {\bibfnamefont {M.}~\bibnamefont
  {Vallisneri}},\ }\href {https://doi.org/10.1103/PhysRevD.77.042001}
  {\bibfield  {journal} {\bibinfo  {journal} {\prd}\ }\textbf {\bibinfo
  {volume} {77}},\ \bibinfo {pages} {042001} (\bibinfo {year}
  {2008})}\BibitemShut {NoStop}%
\bibitem [{\citenamefont {Ho}\ and\ \citenamefont
  {Andersson}(2023)}]{Ho:2023shr}%
  \BibitemOpen
  \bibfield  {author} {\bibinfo {author} {\bibfnamefont {W.~C.~G.}\
  \bibnamefont {Ho}}\ and\ \bibinfo {author} {\bibfnamefont {N.}~\bibnamefont
  {Andersson}},\ }\href {https://doi.org/10.1103/PhysRevD.108.043003}
  {\bibfield  {journal} {\bibinfo  {journal} {\prd}\ }\textbf {\bibinfo
  {volume} {108}},\ \bibinfo {pages} {043003} (\bibinfo {year}
  {2023})}\BibitemShut {NoStop}%
\bibitem [{\citenamefont {Revilla-Pe{\~n}a}\ \emph {et~al.}(2025)\citenamefont
  {Revilla-Pe{\~n}a}, \citenamefont {Bondarescu}, \citenamefont {Lundgren},\
  and\ \citenamefont {Miralda-Escud{\'e}}}]{Revilla-Pena:2025ztm}%
  \BibitemOpen
  \bibfield  {author} {\bibinfo {author} {\bibfnamefont {A.}~\bibnamefont
  {Revilla-Pe{\~n}a}}, \bibinfo {author} {\bibfnamefont {R.}~\bibnamefont
  {Bondarescu}}, \bibinfo {author} {\bibfnamefont {A.~P.}\ \bibnamefont
  {Lundgren}},\ and\ \bibinfo {author} {\bibfnamefont {J.}~\bibnamefont
  {Miralda-Escud{\'e}}},\ }\href {https://doi.org/10.1103/6gm3-q143} {\bibfield
   {journal} {\bibinfo  {journal} {\prd}\ }\textbf {\bibinfo {volume} {112}},\
  \bibinfo {pages} {103011} (\bibinfo {year} {2025})}\BibitemShut {NoStop}%
\bibitem [{\citenamefont {Poisson}(2020)}]{Poisson:2020eki}%
  \BibitemOpen
  \bibfield  {author} {\bibinfo {author} {\bibfnamefont {E.}~\bibnamefont
  {Poisson}},\ }\href {https://doi.org/10.1103/PhysRevD.101.104028} {\bibfield
  {journal} {\bibinfo  {journal} {\prd}\ }\textbf {\bibinfo {volume} {101}},\
  \bibinfo {pages} {104028} (\bibinfo {year} {2020})}\BibitemShut {NoStop}%
\bibitem [{\citenamefont {Pratten}\ \emph
  {et~al.}(2020{\natexlab{b}})\citenamefont {Pratten}, \citenamefont {Husa},
  \citenamefont {Garcia-Quiros}, \citenamefont {Colleoni}, \citenamefont
  {Ramos-Buades}, \citenamefont {Estelles},\ and\ \citenamefont
  {Jaume}}]{Pratten:2020fqn}%
  \BibitemOpen
  \bibfield  {author} {\bibinfo {author} {\bibfnamefont {G.}~\bibnamefont
  {Pratten}}, \bibinfo {author} {\bibfnamefont {S.}~\bibnamefont {Husa}},
  \bibinfo {author} {\bibfnamefont {C.}~\bibnamefont {Garcia-Quiros}}, \bibinfo
  {author} {\bibfnamefont {M.}~\bibnamefont {Colleoni}}, \bibinfo {author}
  {\bibfnamefont {A.}~\bibnamefont {Ramos-Buades}}, \bibinfo {author}
  {\bibfnamefont {H.}~\bibnamefont {Estelles}},\ and\ \bibinfo {author}
  {\bibfnamefont {R.}~\bibnamefont {Jaume}},\ }\href
  {https://doi.org/10.1103/PhysRevD.102.064001} {\bibfield  {journal} {\bibinfo
   {journal} {\prd}\ }\textbf {\bibinfo {volume} {102}},\ \bibinfo {pages}
  {064001} (\bibinfo {year} {2020}{\natexlab{b}})}\BibitemShut {NoStop}%
\bibitem [{\citenamefont {{LIGO Scientific Collaboration}}\ \emph
  {et~al.}(2018)\citenamefont {{LIGO Scientific Collaboration}}, \citenamefont
  {{Virgo Collaboration}},\ and\ \citenamefont {{KAGRA
  Collaboration}}}]{lalsuite}%
  \BibitemOpen
  \bibfield  {author} {\bibinfo {author} {\bibnamefont {{LIGO Scientific
  Collaboration}}}, \bibinfo {author} {\bibnamefont {{Virgo Collaboration}}},\
  and\ \bibinfo {author} {\bibnamefont {{KAGRA Collaboration}}},\ }\href
  {https://doi.org/10.7935/GT1W-FZ16} {\bibinfo {title} {{LVK} {A}lgorithm
  {L}ibrary - {LALS}uite}},\ \bibinfo {howpublished} {Free software (GPL)}
  (\bibinfo {year} {2018})\BibitemShut {NoStop}%
\bibitem [{\citenamefont {O'Shaughnessy}\ \emph {et~al.}(2008)\citenamefont
  {O'Shaughnessy}, \citenamefont {Kim}, \citenamefont {Kalogera},\ and\
  \citenamefont {Belczynski}}]{OShaughnessy:2006uzj}%
  \BibitemOpen
  \bibfield  {author} {\bibinfo {author} {\bibfnamefont {R.~W.}\ \bibnamefont
  {O'Shaughnessy}}, \bibinfo {author} {\bibfnamefont {C.}~\bibnamefont {Kim}},
  \bibinfo {author} {\bibfnamefont {V.}~\bibnamefont {Kalogera}},\ and\
  \bibinfo {author} {\bibfnamefont {K.}~\bibnamefont {Belczynski}},\ }\href
  {https://doi.org/10.1086/523620} {\bibfield  {journal} {\bibinfo  {journal}
  {\apj}\ }\textbf {\bibinfo {volume} {672}},\ \bibinfo {pages} {479} (\bibinfo
  {year} {2008})}\BibitemShut {NoStop}%
\bibitem [{\citenamefont {{McDermott}}\ \emph {et~al.}(1988)\citenamefont
  {{McDermott}}, \citenamefont {{van Horn}},\ and\ \citenamefont
  {{Hansen}}}]{1988ApJ...325..725M}%
  \BibitemOpen
  \bibfield  {author} {\bibinfo {author} {\bibfnamefont {P.~N.}\ \bibnamefont
  {{McDermott}}}, \bibinfo {author} {\bibfnamefont {H.~M.}\ \bibnamefont {{van
  Horn}}},\ and\ \bibinfo {author} {\bibfnamefont {C.~J.}\ \bibnamefont
  {{Hansen}}},\ }\href {https://doi.org/10.1086/166044} {\bibfield  {journal}
  {\bibinfo  {journal} {\apj}\ }\textbf {\bibinfo {volume} {325}},\ \bibinfo
  {pages} {725} (\bibinfo {year} {1988})}\BibitemShut {NoStop}%
\bibitem [{\citenamefont {Koehn}\ \emph {et~al.}(2025)\citenamefont {Koehn}
  \emph {et~al.}}]{Koehn:2024set}%
  \BibitemOpen
  \bibfield  {author} {\bibinfo {author} {\bibfnamefont {H.}~\bibnamefont
  {Koehn}} \emph {et~al.},\ }\href {https://doi.org/10.1103/PhysRevX.15.021014}
  {\bibfield  {journal} {\bibinfo  {journal} {\prx}\ }\textbf {\bibinfo
  {volume} {15}},\ \bibinfo {pages} {021014} (\bibinfo {year}
  {2025})}\BibitemShut {NoStop}%
\bibitem [{\citenamefont {Iorio}\ \emph {et~al.}(2023)\citenamefont {Iorio}
  \emph {et~al.}}]{Iorio:2022sgz}%
  \BibitemOpen
  \bibfield  {author} {\bibinfo {author} {\bibfnamefont {G.}~\bibnamefont
  {Iorio}} \emph {et~al.},\ }\href {https://doi.org/10.1093/mnras/stad1630}
  {\bibfield  {journal} {\bibinfo  {journal} {\mnras}\ }\textbf {\bibinfo
  {volume} {524}},\ \bibinfo {pages} {426} (\bibinfo {year}
  {2023})}\BibitemShut {NoStop}%
\bibitem [{\citenamefont {Aghanim}\ \emph {et~al.}(2020)\citenamefont {Aghanim}
  \emph {et~al.}}]{Planck:2018vyg}%
  \BibitemOpen
  \bibfield  {author} {\bibinfo {author} {\bibfnamefont {N.}~\bibnamefont
  {Aghanim}} \emph {et~al.} (\bibinfo {collaboration} {Planck}),\ }\href
  {https://doi.org/10.1051/0004-6361/201833910} {\bibfield  {journal} {\bibinfo
   {journal} {\aap}\ }\textbf {\bibinfo {volume} {641}},\ \bibinfo {pages} {A6}
  (\bibinfo {year} {2020})},\ \bibinfo {note} {[Erratum: \aap\ \textbf{652}, C4
  (2021)]}\BibitemShut {NoStop}%
\bibitem [{\citenamefont {Hild}\ \emph {et~al.}(2011)\citenamefont {Hild} \emph
  {et~al.}}]{Hild:2010id}%
  \BibitemOpen
  \bibfield  {author} {\bibinfo {author} {\bibfnamefont {S.}~\bibnamefont
  {Hild}} \emph {et~al.},\ }\href
  {https://doi.org/10.1088/0264-9381/28/9/094013} {\bibfield  {journal}
  {\bibinfo  {journal} {\cqg}\ }\textbf {\bibinfo {volume} {28}},\ \bibinfo
  {pages} {094013} (\bibinfo {year} {2011})}\BibitemShut {NoStop}%
\bibitem [{\citenamefont {Speagle}(2020)}]{Speagle:2019ivv}%
  \BibitemOpen
  \bibfield  {author} {\bibinfo {author} {\bibfnamefont {J.~S.}\ \bibnamefont
  {Speagle}},\ }\href {https://doi.org/10.1093/mnras/staa278} {\bibfield
  {journal} {\bibinfo  {journal} {\mnras}\ }\textbf {\bibinfo {volume} {493}},\
  \bibinfo {pages} {3132} (\bibinfo {year} {2020})}\BibitemShut {NoStop}%
\bibitem [{\citenamefont {Ashton}\ \emph {et~al.}(2019)\citenamefont {Ashton}
  \emph {et~al.}}]{Ashton:2018jfp}%
  \BibitemOpen
  \bibfield  {author} {\bibinfo {author} {\bibfnamefont {G.}~\bibnamefont
  {Ashton}} \emph {et~al.},\ }\href {https://doi.org/10.3847/1538-4365/ab06fc}
  {\bibfield  {journal} {\bibinfo  {journal} {\apjss}\ }\textbf {\bibinfo
  {volume} {241}},\ \bibinfo {pages} {27} (\bibinfo {year} {2019})}\BibitemShut
  {NoStop}%
\bibitem [{\citenamefont {Zackay}\ \emph {et~al.}(2018)\citenamefont {Zackay},
  \citenamefont {Dai},\ and\ \citenamefont {Venumadhav}}]{Zackay:2018qdy}%
  \BibitemOpen
  \bibfield  {author} {\bibinfo {author} {\bibfnamefont {B.}~\bibnamefont
  {Zackay}}, \bibinfo {author} {\bibfnamefont {L.}~\bibnamefont {Dai}},\ and\
  \bibinfo {author} {\bibfnamefont {T.}~\bibnamefont {Venumadhav}},\ }\Eprint
  {https://arxiv.org/abs/1806.08792} {arXiv:1806.08792 [astro-ph.IM]}
  (\bibinfo {year} {2018})\BibitemShut {NoStop}%
\bibitem [{\citenamefont {Narola}\ \emph {et~al.}(2024)\citenamefont {Narola},
  \citenamefont {Janquart}, \citenamefont {Meijer}, \citenamefont {Haris},\
  and\ \citenamefont {Van Den~Broeck}}]{Narola:2023men}%
  \BibitemOpen
  \bibfield  {author} {\bibinfo {author} {\bibfnamefont {H.}~\bibnamefont
  {Narola}}, \bibinfo {author} {\bibfnamefont {J.}~\bibnamefont {Janquart}},
  \bibinfo {author} {\bibfnamefont {Q.}~\bibnamefont {Meijer}}, \bibinfo
  {author} {\bibfnamefont {K.}~\bibnamefont {Haris}},\ and\ \bibinfo {author}
  {\bibfnamefont {C.}~\bibnamefont {Van Den~Broeck}},\ }\href
  {https://doi.org/10.1103/PhysRevD.110.084085} {\bibfield  {journal} {\bibinfo
   {journal} {\prd}\ }\textbf {\bibinfo {volume} {110}},\ \bibinfo {pages}
  {084085} (\bibinfo {year} {2024})}\BibitemShut {NoStop}%
\bibitem [{\citenamefont {Leslie}\ \emph {et~al.}(2021)\citenamefont {Leslie},
  \citenamefont {Dai},\ and\ \citenamefont {Pratten}}]{Leslie:2021ssu}%
  \BibitemOpen
  \bibfield  {author} {\bibinfo {author} {\bibfnamefont {N.}~\bibnamefont
  {Leslie}}, \bibinfo {author} {\bibfnamefont {L.}~\bibnamefont {Dai}},\ and\
  \bibinfo {author} {\bibfnamefont {G.}~\bibnamefont {Pratten}},\ }\href
  {https://doi.org/10.1103/PhysRevD.104.123030} {\bibfield  {journal} {\bibinfo
   {journal} {\prd}\ }\textbf {\bibinfo {volume} {104}},\ \bibinfo {pages}
  {123030} (\bibinfo {year} {2021})}\BibitemShut {NoStop}%
\bibitem [{\citenamefont {Nitz}\ and\ \citenamefont
  {Dal~Canton}(2021)}]{Nitz:2021pbr}%
  \BibitemOpen
  \bibfield  {author} {\bibinfo {author} {\bibfnamefont {A.~H.}\ \bibnamefont
  {Nitz}}\ and\ \bibinfo {author} {\bibfnamefont {T.}~\bibnamefont
  {Dal~Canton}},\ }\href {https://doi.org/10.3847/2041-8213/ac1a75} {\bibfield
  {journal} {\bibinfo  {journal} {\apj}\ }\textbf {\bibinfo {volume} {917}},\
  \bibinfo {pages} {L27} (\bibinfo {year} {2021})}\BibitemShut {NoStop}%
\bibitem [{\citenamefont {Veitch}\ and\ \citenamefont
  {Vecchio}(2010)}]{Veitch:2009hd}%
  \BibitemOpen
  \bibfield  {author} {\bibinfo {author} {\bibfnamefont {J.}~\bibnamefont
  {Veitch}}\ and\ \bibinfo {author} {\bibfnamefont {A.}~\bibnamefont
  {Vecchio}},\ }\href {https://doi.org/10.1103/PhysRevD.81.062003} {\bibfield
  {journal} {\bibinfo  {journal} {\prd}\ }\textbf {\bibinfo {volume} {81}},\
  \bibinfo {pages} {062003} (\bibinfo {year} {2010})}\BibitemShut {NoStop}%
\bibitem [{\citenamefont {{Reisenegger}}\ and\ \citenamefont
  {{Goldreich}}(1992)}]{1992ApJ...395..240R}%
  \BibitemOpen
  \bibfield  {author} {\bibinfo {author} {\bibfnamefont {A.}~\bibnamefont
  {{Reisenegger}}}\ and\ \bibinfo {author} {\bibfnamefont {P.}~\bibnamefont
  {{Goldreich}}},\ }\href {https://doi.org/10.1086/171645} {\bibfield
  {journal} {\bibinfo  {journal} {\apj}\ }\textbf {\bibinfo {volume} {395}},\
  \bibinfo {pages} {240} (\bibinfo {year} {1992})}\BibitemShut {NoStop}%
\bibitem [{\citenamefont {Counsell}\ \emph {et~al.}(2024)\citenamefont
  {Counsell}, \citenamefont {Gittins}, \citenamefont {Andersson},\ and\
  \citenamefont {Pnigouras}}]{Counsell:2024pua}%
  \BibitemOpen
  \bibfield  {author} {\bibinfo {author} {\bibfnamefont {R.}~\bibnamefont
  {Counsell}}, \bibinfo {author} {\bibfnamefont {F.}~\bibnamefont {Gittins}},
  \bibinfo {author} {\bibfnamefont {N.}~\bibnamefont {Andersson}},\ and\
  \bibinfo {author} {\bibfnamefont {P.}~\bibnamefont {Pnigouras}},\ }\href
  {https://doi.org/10.1093/mnras/stae2721} {\bibfield  {journal} {\bibinfo
  {journal} {\mnras}\ }\textbf {\bibinfo {volume} {536}},\ \bibinfo {pages}
  {1967} (\bibinfo {year} {2024})}\BibitemShut {NoStop}%
\bibitem [{\citenamefont {Lockitch}\ and\ \citenamefont
  {Friedman}(1999)}]{Lockitch:1998nq}%
  \BibitemOpen
  \bibfield  {author} {\bibinfo {author} {\bibfnamefont {K.~H.}\ \bibnamefont
  {Lockitch}}\ and\ \bibinfo {author} {\bibfnamefont {J.~L.}\ \bibnamefont
  {Friedman}},\ }\href {https://doi.org/10.1086/307580} {\bibfield  {journal}
  {\bibinfo  {journal} {\apj}\ }\textbf {\bibinfo {volume} {521}},\ \bibinfo
  {pages} {764} (\bibinfo {year} {1999})}\BibitemShut {NoStop}%
\bibitem [{\citenamefont {Tsang}\ \emph {et~al.}(2012)\citenamefont {Tsang},
  \citenamefont {Read}, \citenamefont {Hinderer}, \citenamefont {Piro},\ and\
  \citenamefont {Bondarescu}}]{Tsang:2011ad}%
  \BibitemOpen
  \bibfield  {author} {\bibinfo {author} {\bibfnamefont {D.}~\bibnamefont
  {Tsang}}, \bibinfo {author} {\bibfnamefont {J.~S.}\ \bibnamefont {Read}},
  \bibinfo {author} {\bibfnamefont {T.}~\bibnamefont {Hinderer}}, \bibinfo
  {author} {\bibfnamefont {A.~L.}\ \bibnamefont {Piro}},\ and\ \bibinfo
  {author} {\bibfnamefont {R.}~\bibnamefont {Bondarescu}},\ }\href
  {https://doi.org/10.1103/PhysRevLett.108.011102} {\bibfield  {journal}
  {\bibinfo  {journal} {\prl}\ }\textbf {\bibinfo {volume} {108}},\ \bibinfo
  {pages} {011102} (\bibinfo {year} {2012})}\BibitemShut {NoStop}%
\bibitem [{\citenamefont {Gao}\ \emph {et~al.}(2025)\citenamefont {Gao},
  \citenamefont {Kuan}, \citenamefont {Xia}, \citenamefont {Silva},\ and\
  \citenamefont {Shibata}}]{Gao:2025aqo}%
  \BibitemOpen
  \bibfield  {author} {\bibinfo {author} {\bibfnamefont {Y.}~\bibnamefont
  {Gao}}, \bibinfo {author} {\bibfnamefont {H.-J.}\ \bibnamefont {Kuan}},
  \bibinfo {author} {\bibfnamefont {C.-J.}\ \bibnamefont {Xia}}, \bibinfo
  {author} {\bibfnamefont {H.~O.}\ \bibnamefont {Silva}},\ and\ \bibinfo
  {author} {\bibfnamefont {M.}~\bibnamefont {Shibata}},\ }\href
  {https://doi.org/10.1103/nbk7-8kts} {\bibfield  {journal} {\bibinfo
  {journal} {\prd}\ }\textbf {\bibinfo {volume} {112}},\ \bibinfo {pages}
  {123006} (\bibinfo {year} {2025})}\BibitemShut {NoStop}%
\bibitem [{\citenamefont {Miao}\ \emph {et~al.}(2024)\citenamefont {Miao},
  \citenamefont {Zhou},\ and\ \citenamefont {Li}}]{Miao:2023jqe}%
  \BibitemOpen
  \bibfield  {author} {\bibinfo {author} {\bibfnamefont {Z.}~\bibnamefont
  {Miao}}, \bibinfo {author} {\bibfnamefont {E.}~\bibnamefont {Zhou}},\ and\
  \bibinfo {author} {\bibfnamefont {A.}~\bibnamefont {Li}},\ }\href
  {https://doi.org/10.3847/1538-4357/ad27cd} {\bibfield  {journal} {\bibinfo
  {journal} {\apj}\ }\textbf {\bibinfo {volume} {964}},\ \bibinfo {pages} {31}
  (\bibinfo {year} {2024})}\BibitemShut {NoStop}%
\bibitem [{\citenamefont {Counsell}\ \emph {et~al.}(2025)\citenamefont
  {Counsell}, \citenamefont {Gittins}, \citenamefont {Andersson},\ and\
  \citenamefont {Tews}}]{Counsell:2025hcv}%
  \BibitemOpen
  \bibfield  {author} {\bibinfo {author} {\bibfnamefont {A.~R.}\ \bibnamefont
  {Counsell}}, \bibinfo {author} {\bibfnamefont {F.}~\bibnamefont {Gittins}},
  \bibinfo {author} {\bibfnamefont {N.}~\bibnamefont {Andersson}},\ and\
  \bibinfo {author} {\bibfnamefont {I.}~\bibnamefont {Tews}},\ }\href
  {https://doi.org/10.1103/8hvq-6dy7} {\bibfield  {journal} {\bibinfo
  {journal} {\prl}\ }\textbf {\bibinfo {volume} {135}},\ \bibinfo {pages}
  {081402} (\bibinfo {year} {2025})}\BibitemShut {NoStop}%
\bibitem [{\citenamefont {Pereira}\ \emph {et~al.}(2025)\citenamefont
  {Pereira}, \citenamefont {Tonetto}, \citenamefont {Bejger}, \citenamefont
  {Zdunik},\ and\ \citenamefont {Haensel}}]{Pereira:2025xsi}%
  \BibitemOpen
  \bibfield  {author} {\bibinfo {author} {\bibfnamefont {J.~P.}\ \bibnamefont
  {Pereira}}, \bibinfo {author} {\bibfnamefont {L.}~\bibnamefont {Tonetto}},
  \bibinfo {author} {\bibfnamefont {M.}~\bibnamefont {Bejger}}, \bibinfo
  {author} {\bibfnamefont {J.~L.}\ \bibnamefont {Zdunik}},\ and\ \bibinfo
  {author} {\bibfnamefont {P.}~\bibnamefont {Haensel}},\ }\href
  {https://doi.org/10.1103/k7l9-hw8g} {\bibfield  {journal} {\bibinfo
  {journal} {\prl}\ }\textbf {\bibinfo {volume} {135}},\ \bibinfo {pages}
  {231401} (\bibinfo {year} {2025})}\BibitemShut {NoStop}%
\bibitem [{\citenamefont {Yu}\ \emph {et~al.}(2024)\citenamefont {Yu},
  \citenamefont {Arras},\ and\ \citenamefont {Weinberg}}]{Yu:2024uxt}%
  \BibitemOpen
  \bibfield  {author} {\bibinfo {author} {\bibfnamefont {H.}~\bibnamefont
  {Yu}}, \bibinfo {author} {\bibfnamefont {P.}~\bibnamefont {Arras}},\ and\
  \bibinfo {author} {\bibfnamefont {N.~N.}\ \bibnamefont {Weinberg}},\ }\href
  {https://doi.org/10.1103/PhysRevD.110.024039} {\bibfield  {journal} {\bibinfo
   {journal} {\prd}\ }\textbf {\bibinfo {volume} {110}},\ \bibinfo {pages}
  {024039} (\bibinfo {year} {2024})}\BibitemShut {NoStop}%
\bibitem [{\citenamefont {Pnigouras}\ \emph {et~al.}(2025)\citenamefont
  {Pnigouras}, \citenamefont {Andersson}, \citenamefont {Gittins},\ and\
  \citenamefont {Counsell}}]{Pnigouras:2025muo}%
  \BibitemOpen
  \bibfield  {author} {\bibinfo {author} {\bibfnamefont {P.}~\bibnamefont
  {Pnigouras}}, \bibinfo {author} {\bibfnamefont {N.}~\bibnamefont
  {Andersson}}, \bibinfo {author} {\bibfnamefont {F.}~\bibnamefont {Gittins}},\
  and\ \bibinfo {author} {\bibfnamefont {A.~R.}\ \bibnamefont {Counsell}},\
  }\href {https://doi.org/10.1093/mnras/staf1285} {\bibfield  {journal}
  {\bibinfo  {journal} {\mnras}\ }\textbf {\bibinfo {volume} {542}},\ \bibinfo
  {pages} {1375} (\bibinfo {year} {2025})}\BibitemShut {NoStop}%
\bibitem [{\citenamefont {Gittins}\ \emph {et~al.}(2026)\citenamefont
  {Gittins}, \citenamefont {Narola}, \citenamefont {Wouters}, \citenamefont
  {Pang}, \citenamefont {Hinderer},\ and\ \citenamefont {Van
  Den~Broeck}}]{zenodo}%
  \BibitemOpen
  \bibfield  {author} {\bibinfo {author} {\bibfnamefont {F.}~\bibnamefont
  {Gittins}}, \bibinfo {author} {\bibfnamefont {H.}~\bibnamefont {Narola}},
  \bibinfo {author} {\bibfnamefont {T.}~\bibnamefont {Wouters}}, \bibinfo
  {author} {\bibfnamefont {P.~T.~H.}\ \bibnamefont {Pang}}, \bibinfo {author}
  {\bibfnamefont {T.}~\bibnamefont {Hinderer}},\ and\ \bibinfo {author}
  {\bibfnamefont {C.}~\bibnamefont {Van Den~Broeck}},\ }\href
  {https://doi.org/10.5281/zenodo.20441757} {\bibinfo {title} {{Detecting Tidal
  Resonances in Binary Neutron Stars: Data Products}}} (\bibinfo {year}
  {2026})\BibitemShut {NoStop}%
\end{thebibliography}%

\end{document}